\documentclass[%
 reprint,
superscriptaddress,
 amsmath,amssymb,
 aps,
]{revtex4-2}

\usepackage{graphicx}
\usepackage{dcolumn}
\usepackage{bm}

\usepackage{color}      

\newcommand{\microm}{\upmu\textrm{m}}
\newcommand{\kbt}{k_\textrm{B}T}

\usepackage{textgreek}
\usepackage{upgreek}
\usepackage{hyperref}

\begin{document}

\preprint{APS/123-QED}

\title{Erythrocyte sedimentation:\\Fracture and collapse of a high-volume-fraction soft-colloid gel }

\author{Alexis Darras}
\email{alexis.charles.darras@gmail.com}
\affiliation{Experimental Physics, Saarland University, 66123 Saarbruecken, Germany
}
\author{Anil Kumar Dasanna}
\affiliation{Theoretical Physics of Living Matter, Institute of Biological Information Processing and Institute for Advanced Simulation, Forschungszentrum Jülich, 52425 Jülich, Germany}
\author{Thomas John}%
\affiliation{Experimental Physics, Saarland University, 66123 Saarbruecken, Germany
}
\author{Gerhard Gompper}
\affiliation{Theoretical Physics of Living Matter, Institute of Biological Information Processing and Institute for Advanced Simulation, Forschungszentrum Jülich, 52425 Jülich, Germany}
\author{Lars Kaestner}
\affiliation{Experimental Physics, Saarland University, 66123 Saarbruecken, Germany
}
\affiliation{Theoretical Medicine and Biosciences, Saarland University, 66424 Homburg, Germany}
\author{Dmitry A. Fedosov}
\affiliation{Theoretical Physics of Living Matter, Institute of Biological Information Processing and Institute for Advanced Simulation, Forschungszentrum Jülich, 52425 Jülich, Germany}
\author{Christian Wagner}
\affiliation{Experimental Physics, Saarland University, 66123 Saarbruecken, Germany
}
\affiliation{Physics and Materials Science Research Unit, University of Luxembourg, Luxembourg City, Luxembourg}

\date{\today}

\begin{abstract}
The erythrocyte sedimentation rate is one of the oldest medical diagnostic methods whose physical mechanisms remain debatable up to date. Using both light microscopy and mesoscale cell-level simulations, we show that erythrocytes form a soft-colloid gel. Furthermore, the high volume fraction of erythrocytes, their deformability, and weak attraction lead to unusual properties of this gel. A theoretical model for the gravitational collapse is developed, whose predictions are  in agreement with detailed macroscopic measurements of the interface velocity.
\end{abstract}

\maketitle


\begin{figure*}[!hbt]
\includegraphics[width=0.95\textwidth]{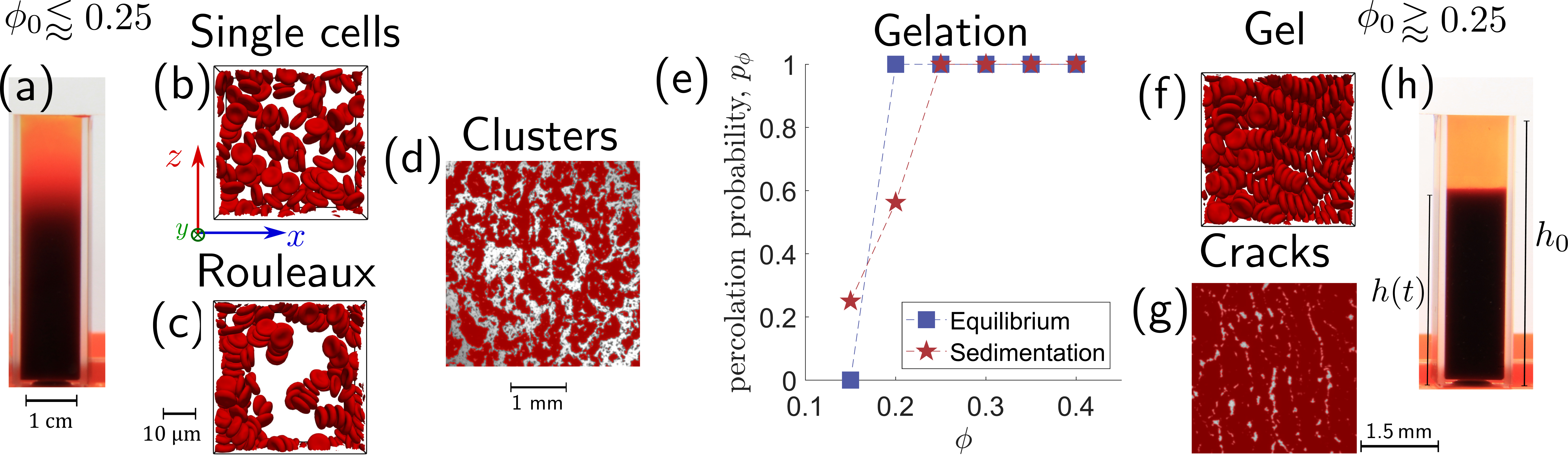}

\caption{\label{Regimes}Possible regimes of erythrocyte sedimentation. (a) Experimental photograph of a low volume fraction ($\phi_0=0.2$) erythrocyte suspension, after $20~\mathrm{min}$ at rest in a cuvette with quadratic cross-section. A diffuse interface hints to a sedimentation of separate aggregates. (b) Snapshot from simulation: well-dispersed suspension of erythrocytes for $\phi=0.25$ and no attraction force between the cells. (c) Snapshot from simulation: formation of rouleaux structures at $\phi=0.25$ and physiological attraction between the cells ($\epsilon=2.5~\kbt$, see the joint paper for details \cite{JointPaper}). (d) False-color photograph of erythrocytes clusters in a $150~\mathrm{\upmu m}$ thick container, at volume fraction $\phi=0.20$ (see also Supplemental Movie S1 \cite{SuppMat}). (e) Percolation probability $p_\phi$ is plotted against hematocrit $\phi$ (simulations have a constant hematocrit due to periodic boundary conditions on $z$).  The probability $p_\phi$ is defined as the fraction of snapshots in which the largest cluster is percolating in both $x$ and $y$ (periodic) directions. The data is shown for both sedimentation and equilibrium. (f) A gel-like network for $\phi=0.4$. All simulation snapshots have the same scale. (g) False-color photograph of the cracks for blood sedimentation in a $150~\mathrm{\upmu m}$ thick container, at volume fraction $\phi=0.45$ (see also Supplemental Movie S2 \cite{SuppMat}). (h) Erythrocyte suspension with physiological volume fraction ($\phi_0=0.45$), after $2~\mathrm{hours}$ at rest. A sharp interface, indicating a gel-like collapse behavior, is observed. }
\end{figure*}

The erythrocyte sedimentation rate (ESR) is one of the oldest medical diagnostic tools with its origin rooting in ancient Greek times \cite{kushner1988acute}. It was popularized in the late 19th/early 20th century as a non-specific test to diagnose and monitor inflammation \cite{grzybowski2011edmund,bedell1985,tishkowski2020erythrocyte}. However, recent research has shown that it might also be a good and cheap biomarker to detect abnormal, deformed erythrocytes (or red blood cells (RBCs)) \cite{darras2021acanthocyte}. Despite its frequent use in medicine, physical mechanisms that govern the sedimentation of erythrocytes did not advance much during the last century. It is generally accepted that erythrocytes agglomerate into separate aggregates whose sedimentation is well described by a Stokes-like law \cite{taye2020sedimentation,smallwood1985physics,puccini1977erythrocyte,dorrington1983erythrocyte,sousa2018validation}. However, the corresponding physical models for erythrocyte sedimentation are often semi-empirical, and provide at best a qualitative description \cite{puccini1977erythrocyte,dorrington1983erythrocyte,sousa2018validation}. 

Several decades of research on the sedimentation of colloidal suspensions have shown that many suspensions of attractive particles form a colloidal gel, which sediments very differently from a collection of separate aggregates. For instance, the sedimentation of such gels is characterized by a delayed, sudden collapse, sometimes associated with the emergence of cracks within the gel structure \cite{derec2003rapid,bartlett2012sudden}. These gels also exhibit a sharp interface between the free liquid phase and the gel of particles \cite{allain1995aggregation,manley2005gravitational,padmanabhan2018,rouwhorst2020nonequilibrium}. In this letter, we show that this colloidal-gel interpretation provides a consistent and quantitative description of the sedimentation of erythrocytes at physiological volume fractions. In particular, we show that erythrocytes form a soft-colloid gel whose sedimentation is well described by a theoretical description based on the collapse of gel-like suspensions. Furthermore, complementary simulations demonstrate that erythrocyte flexibility and high volume fraction are important parameters that determine the gel-like behavior of erythrocyte suspension (or blood).

Despite the fact that ESR is often explained as the sedimentation of separate aggregates, several current studies already provide some arguments supporting the idea of gel-like sedimentation: (i) The formation of distinct aggregates of RBCs should result in a strongly polydisperse suspension \cite{ami2001parameters}. Suspensions with polydisperse aggregates form a diffuse interface during sedimentation \cite{xue2003modeling,watson2005sedimentation,selim1983sedimentation}, contrary to what is observed for erythrocyte sedimentation at physiological hematocrit (volume fraction of RBCs) values of $0.35<\phi_0<0.55$. (ii) Due to the low shear stress during their sedimentation, the aggregation process of erythrocytes is nearly irreversible. Such an aggregation on the time scale of minutes to hours leads to a percolating (or space-filling) network of RBCs \cite{pribush2000dielectric}. This is consistent with a gel-like structure observed for even more dilute suspensions of other attractive colloids \cite{allain1995aggregation}. (iii) The interface position between the dense erythrocyte phase and the free liquid (blood plasma) can actually oscillate or jump in time \cite{tuchin2002dynamic}. (iv) A notable modification in suspension conductivity has been observed for sedimenting erythrocytes, which can be explained by the opening of channels for the liquid phase, i.e. the plasma, within aggregated erythrocyte structures \cite{pribush2010mechanism,pribush2010mechanismII}. Such channels are characteristic for the sedimentation of colloidal gels \cite{derec2003rapid}.

The main reason for the so-far generally accepted picture that erythrocytes sediment as separate aggregates is that the sedimentation velocity increases with the aggregation interaction strength between erythrocytes, which is mainly determined by the concentration of polymers (and in particular fibrinogen) in the plasma \cite{hung1994erythrocyte,holley1999influence}. However, previous models of gravitational collapse of colloidal gels also show that increased attraction increase the rigidity of the network and slow down the sedimentation \cite{manley2005gravitational,channell2000effects,kilfoil2003dynamics,kamp2009universal,dinsmore2006microscopic,pribush2010mechanism1,teece2014gels,lindstrom2012}. Pribush \textit{et al.} \cite{pribush2010mechanism,pribush2010mechanismII} mention the possibility of a transient erythrocyte gel, but claim that such a percolating network must break into separate aggregates during sedimentation due to the induced upward flow of plasma (caused by volume conservation). Furthermore, there is disagreement in how hematocrit $\phi_0$ affects the ESR, where the majority of studies finds that the ESR decreases with $\phi_0$ \cite{poole1952correction,bull1974evaluation,baskurt2011red}, while others support the opposite dependence \cite{pribush2010mechanismII}.

To reconcile these different interpretations, we perform ESR measurements with various volume fractions in combination with 3D hydrodynamic simulations for the analysis of cell-level structure and dynamics. Sedimentation experiments are performed in cuvettes with inner dimensions of $10\times10\times40\,\text{mm}^3$ to minimize wall effects on RBC sedimentation \cite{manley2005gravitational,derec2003rapid}. 

Blood samples were collected from various healthy volunteers with an informed consent \footnote{According to the declaration of Helsinki and the approval by the ethics committee 'Ärztekammer des Saarlandes', ethics votum 51/18.}. 
For $\phi_0 < 0.25$, a diffuse interface between erythrocytes and plasma is observed (Fig.~\ref{Regimes}(a)), while a sharp interface is obtained at higher $\phi$ (Fig.~\ref{Regimes}(g)). This suggests that a transition between the regimes of separated aggregates and soft-colloid gel occurs at $\phi_0\approx0.25$, at otherwise physiological conditions. 

We also performed experiments with a container made of two glass plates separated by a paraffin layer (around $150~\microm$ thick, $2~\mathrm{cm}$ wide and $6~\mathrm{cm}$ high). Those experiments are illustrated in Supplemental Figure S1 \cite{SuppMat}. These containers reveals the separated microscopic aggregates (Supplemental Movie S1 \cite{SuppMat}) for $\phi_0=0.2$ and the apparition of channels in a cohesive gel for $\phi_0=0.45$ (Movie S2 \cite{SuppMat}). The Movie S3 \cite{SuppMat} also shows a so-called eruption of the colloidal gel \cite{derec2003rapid}, which were previously hypothesized as the explanation for the fluctuations of the interface \cite{pribush2010mechanism1}.

In order to gain deeper insight into the governing physical mechanisms, we also perform mesoscopic computer simulations of aggregation of deformable RBCs, with biconcave-disc shape at rest, and with short-range inter-cell attractive membrane interactions, combined with fluid modelled by Smoothed Dissipative Particle Dynamics (SDPD) \cite{fedosov_BMMB2014, Mueller_SDPD2015}. The simulation domain is cubic with a side length $L=50~\microm$ and has periodic boundary conditions in all three directions. Constant volume fractions in the range $\phi\in\left[0.15;0.4\right]$ were probed, meaning between $\sim 200$ and $500$ cells are simulated. We perform simulations in thermal equilibrium, and in a gravitational field which induces upward liquid flow. In both cases, the simulations show aggregated gel-like networks spanning the whole simulation domain, see Fig.~\ref{Regimes}(d).  To check if the cluster is percolating, for our case of periodic boundary conditions, we employ periodic images of the box in the $x$- and $y$-directions (excluding the sedimentation direction, i.e the $z$-direction) and measure the cluster size in these two directions. If the cluster is percolating, the cluster size is equal to $2L$. If the cluster is not percolating, the cluster size becomes smaller than $L$ (but is often close to $L$). The percolation probability is then defined as the fraction of times the largest cluster, which contains a large fraction of all cells, is percolating in both $x$- and $y$-directions. In Fig.~\ref{Regimes}(g), we plot the percolation probability $p_\phi$ for different hematocrit values for both equilibrium and sedimentation. For smaller hematocrit values, the largest cluster breaks into smaller clusters representing the ``fluid" state of the suspension. For hematocrits $\phi_0 \ge 0.30$, the largest cluster is always percolating, in agreement with experimental observations. In the absence of sedimentation, and thus without global upward liquid flow, percolation appears for smaller hematocrit values and the percolation transition with varying volume fraction is sharp.

\begin{figure}[ht!b]
\includegraphics[width=0.42\textwidth]{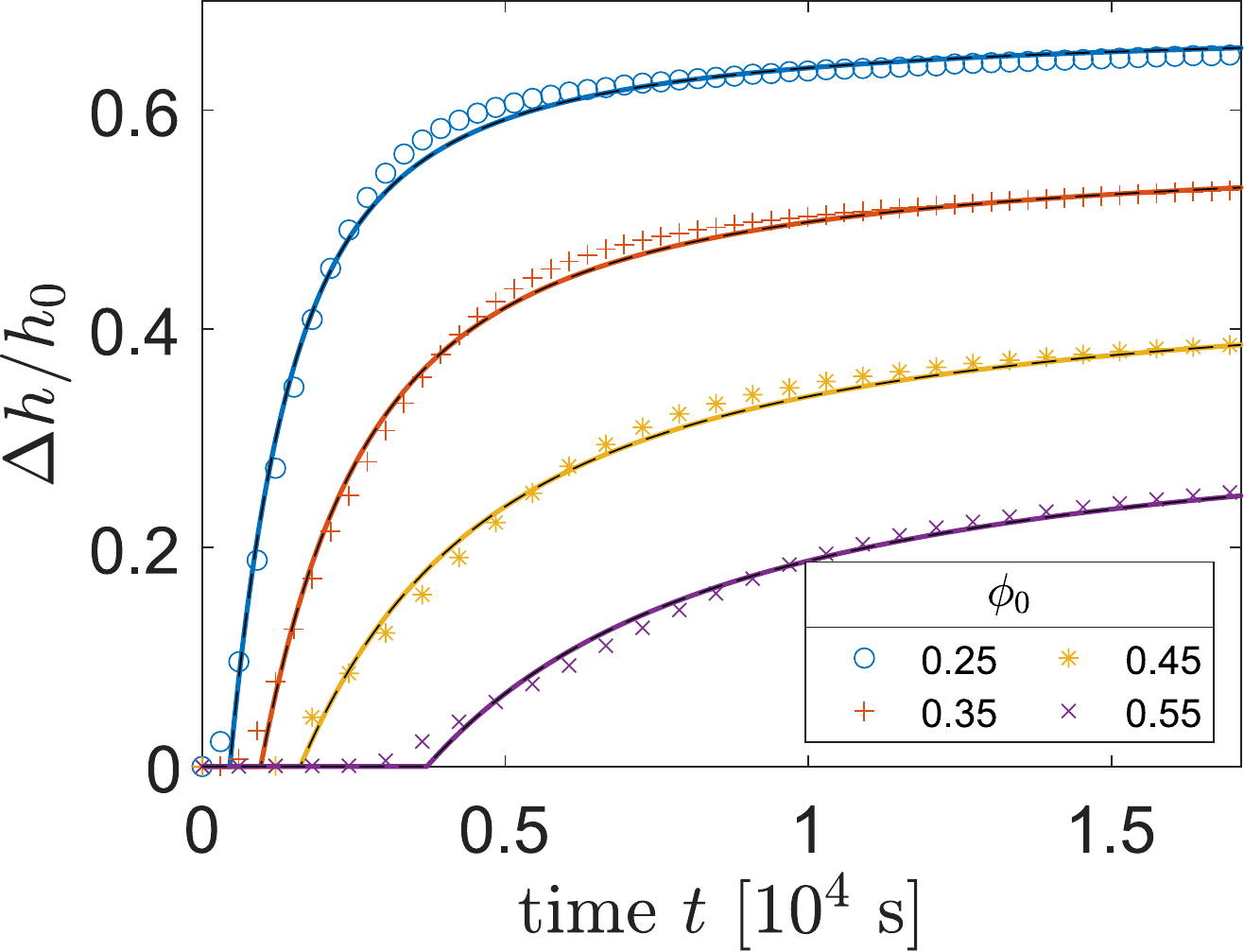}
\caption{\label{PhiCurves}Measurements of relative height reduction $\Delta h/h_0=(h(0)-h(t))/h(0)$ of the dense erythrocyte suspension below the free liquid phase (plasma), for various hematocrits. Note that $\Delta h$ corresponds to 
the height of top plasma layer. The symbols are experimental data, while the curves are fits from our model for $h(t)$ from Eq.~(\ref{VelOne}).}
\end{figure}

\begin{figure*}[tb]
\includegraphics[width=0.85\textwidth]{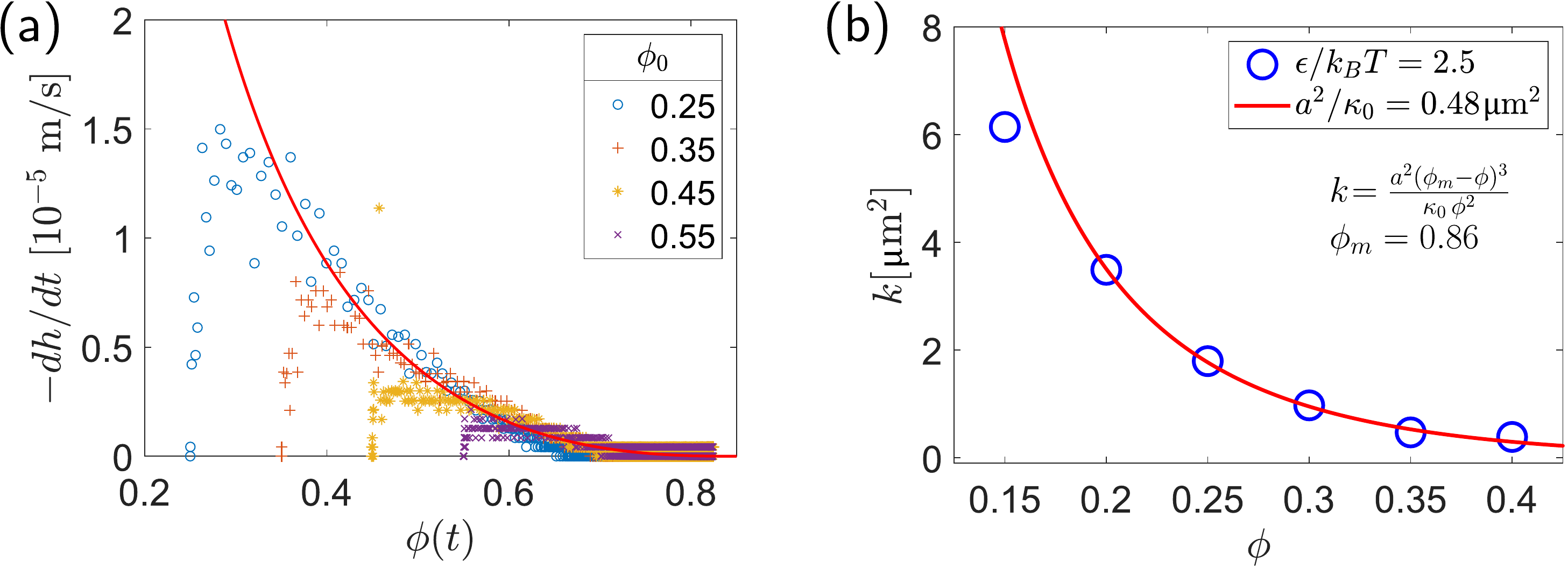}
\caption{\label{PhiVel} Erythrocyte sedimentation for various initial volume fractions. (a) Measurements of velocity of the interface between the erythrocytes gel and the plasma (same experiments as in Fig.~\ref{PhiCurves}). The symbols are experimental data. The average velocity between two measurement points is $dh/dt$, and $\phi(t)=\phi_0h_0/h(t)$. The solid curve is obtained from Eq.~(\ref{VelOne}) using average values of $\gamma=0.42\pm0.06$ and $\phi_m=0.86\pm0.04$ from the entire set of measurements ($16$ samples from seven independent blood drawings). (b) Permeability coefficient from hydrodynamic simulations calculated as $k = (1-\phi)v\eta/\frac{\partial P}{\partial z}$ as a function of hematocrit. The assumed aggregation strength in simulations was $\epsilon=2.5~\kbt$ (see joint paper \cite{JointPaper}).}
\end{figure*}

Further quantitative measurements can be obtained from the experiments by focusing on the interface between the free liquid (plasma) and the dense erythrocytes suspension. Figure~\ref{PhiCurves} shows the variation of this interface position and Fig.~\ref{PhiVel}(a) its velocity. This interface velocity defines the sedimentation velocity, and decreases with increasing hematocrit for volume fractions $\phi_0 \gtrsim 0.25$ (physiological values range from $0.35$ to $0.50$ \cite{chernecky2012laboratory}). An interesting observation is that the delay time of sedimentation increases significantly with increasing $\phi_0$ (see Fig.~\ref{PhiCurves}). This is in agreement with colloidal gels which become more stable with increasing volume fraction of attractive colloids, and thus, require longer times for aging towards an eventual collapse or sedimentation start. For the sedimentation of separate aggregates, the delay time attributed to initial diffusion-limited RBC aggregation should decrease with increasing $\phi_0$, because inter-cell distances decrease and the aggregation should proceed faster \cite{russel1989colloidal}. 

For a quantitative description of the erythrocyte sedimentation process, we develop a semi-empirical model based on the collapse of a colloidal gel. Previous models for sedimenting colloidal gels are not suitable for blood, because they consider colloids with higher attraction, and whose time-dependent volume fraction is assumed to be small such that  $(1-\phi)\gg \phi$ \cite{allain1995aggregation,manley2005gravitational,derec2003rapid}. For blood sedimentation, this assumption is not valid, as $\phi$ initially starts from about $0.4$ and finishes with a value close to unity when sedimentation stops. For simplicity, we assume that the volume fraction $\phi$ of the erythrocyte gel phase is spatially homogeneous but time-dependent. The conservation of erythrocyte volume can be expressed as $h(t)\phi(t)=\phi_0 h_0$, where $h(t)$ is the time-dependent position of the gel surface and $h_0=h(0)$ is the initial height. Assuming a sharp interface between cell-free plasma and erythrocyte gel, total volume conservation allows the connection between $h(t)$ and $v$, the average (upward) velocity of the plasma at the interface, 
\begin{equation}
    (1-\phi) v + \phi \frac{dh}{dt}=0.
    \label{TotVol}
\end{equation}
Considering that pressure gradients which drive upward plasma flow develop due to the gravity force on erythrocytess, we obtain
\begin{equation}
    -(1-\phi)\frac{\partial P}{\partial z}=\Delta \rho g \phi,
    \label{Pressure}
\end{equation}
where $P$ is the pressure, $\Delta \rho$ is the density difference between erythrocyte and plasma, and $g$ is the gravitational acceleration. Note that other possible stresses within erythrocyte gel, such as elastic stresses, are neglected here. 

Flow through a porous material is modelled by Darcy's law,
\begin{equation}
    \left(1-\phi\right)\left(v-\frac{dh}{dt}\right)=-\frac{k}{\eta}\frac{\partial P}{\partial z},
    \label{Darcy}
\end{equation}
where $k$ is the permeability of the medium and $\eta$ is the fluid dynamic viscosity. To approximate the permeability $k$, we use a modified Carman-Kozeny relationship for packed beds of colloids, as proposed by Terzaghi \cite{terzaghi1925} and employed by others \cite{carman1939permeability}
\begin{equation}
    k=\frac{1}{\kappa_0}\frac{a^2 \left(\phi_m-\phi\right)^3}{\phi^2},
    \label{CK}
\end{equation}
where $a$ is the characteristic size of medium pores, $\kappa_0$ is the scaling constant in the Carman-Kozeny relationship that depends on geometric properties of the porous medium \cite{ozgumus2014determination,heijs1995numerical,xu2008developing}, and $\phi_m$ is the maximal volume fraction at which liquid stops flowing through the medium. The parameter $\phi_m$ is a modification introduced initially for compact clay media, because at high volume fractions, the permeability can become zero even for $\phi<1$ \cite{carman1939permeability}. In case of blood, the sedimentation process induces the compaction of erythrocyte packing, which finally stops at a maximal volume fraction $\phi_m<1$. By combining Eqs.~(\ref{TotVol}) to (\ref{CK}), and the elimination of ${\partial P}/{\partial z}$ we obtain an equation for time evolution of the height $h$ of erythrocyte gel as
\begin{equation}
    \frac{dh}{dt}=-\frac{\Delta \rho g a^2}{\kappa_0 \eta}\frac{\left(\phi_m-\phi\right)^3}{\phi\left(1-\phi\right)},
    \label{VelOne}
\end{equation}
where $\phi=\phi_0 h_0/h$ due to the conservation of erythrocyte volume. 
Furthermore, erythrocyte sedimentation exhibits a delayed collapse as shown in Fig.~\ref{PhiCurves}, which is similar to colloidal gels. The origin of sedimentation delay is still under debate, and is likely associated with gel aging and the development of cracks for fluid to flow within the gel \cite{teece2014gels,buscall2009towards,gopalakrishnan2006linking,padmanabhan2018gravitational,bartlett2012sudden}. Such a delay time has already been used in an empirical model for blood sedimentation \cite{hung1994erythrocyte}. In our model, the delay time $t_0$ is introduced as an adjustable parameter, such that $\frac{dh}{dt}=0$ for $t<t_0$ and erythrocyte sedimentation proceeds according to  Eq.~(\ref{VelOne}) for $t>t_0$. 

Even though Eq.~(\ref{VelOne}) does not have a direct analytical solution for $h(t)$, it was used to fit experimental sedimentation results with two additional adjustable parameters: $\phi_m$ and $\gamma = \kappa_0/a^2$ where the latter is related to a characteristic time in this system (see Supplemental Material \cite{SuppMat} and the joint paper \cite{JointPaper} for details).

Several fits using this model are shown in Fig.~\ref{PhiCurves} with a good quantitative agreement between experimental measurements and model predictions for various hematocrits. Supplemental Figure S2 \cite{SuppMat} presents fitting parameters ($\gamma$, $\phi_m$, and $t_0$) for all sedimentation measurements performed. The parameters $\gamma$ and $\phi_m$ display small deviations (respectively upto $15\%$ and $5\%$) without any particular trend, indicating their robustness for the quantification of ESR independently of hematocrit. The observed delay time $t_0$ is the only parameter that strongly depends on $\phi_0$. This is consistent with the idea that the delay in sedimentation is related to the time required for the rearrangement of a gelated network \cite{teece2014gels,gopalakrishnan2006linking}, since larger hematocrit implies stronger connectivity within erythrocyte gel.

Another result of our theoretical description is that the sedimentation velocity is a function of local time-dependent hematocrit. Indeed, Fig.~\ref{PhiVel}(a) compares interface velocities from Eq.~(\ref{VelOne}) and experimental measurements for different $\phi_0$. All collected data collapse onto a single master curve predicted by Eq.~(\ref{VelOne}) with $\gamma = 0.34$ and $\phi_m=0.86$, except a few time points collected below or around the time delay $t_0$. The discrepancies in sedimentation velocity at the very start of sedimentation are likely related to the time required for flow development within an erythrocyte gel. 

The hydrodynamic simulations of sedimentation confirm the relevance of the model. Figure \ref{PhiVel}(b) shows the permeability coefficient $k$, computed through Eq.~(\ref{Darcy}), as a function of hematocrit. These data are fitted with Eq.~(\ref{CK}). The fit closely approximates the simulation data and confirms the validity of Eq.~(\ref{CK}) as constitutive equation.  Furthermore, these hydrodynamic simulations show that for large enough $\phi_0 \gtrsim 0.30$, even though some rearrangements in the gel-like erythrocyte network occur due to sedimenting flow conditions, it still spans the whole simulation domain. Thus, it indicates that for physiological hematocrit values, the initially gelated structure of RBCs likely does not break into separate aggregates during the sedimentation process. 

In summary, we have shown that erythrocyte suspensions sediment as a soft colloidal gel when the physiological range of hematocrits is considered. The developed theoretical model is able to quantitatively describe the behavior of the interface of the erythrocytes gel under conditions relevant for the usual clinical tests. Numerical simulations corroborate the robustness of the theoretical model for erythrocyte sedimentation, and confirm that at physiological hematocrit values, a gel-like network structure of erythrocytes remains even under sedimenting flow conditions. This sedimentation corresponds to a sudden collapse of the gel when it fractures and condenses. The joint paper \cite{JointPaper} presents a systematic investigation of the effect of aggregation strength between erythrocytes on the interface velocity. Our study provides a significant step toward understanding erythrocyte sedimentation employed in classical ESR tests, but also provides a theoretical background for changes in erythrocyte sedimentation related to some diseases (e.g. acanthocytosis), in which erythrocyte properties might be significantly altered \cite{darras2021acanthocyte}.



\bibliography{apssamp}

\end{document}


\preprint{APS/123-QED}

\title{Supplemental Material for\\Erythrocyte sedimentation:\\fracture and collapse of a high-volume-fraction soft-colloid gel}

\author{Alexis Darras}
\email{alexis.charles.darras@gmail.com}
\affiliation{Experimental Physics, Saarland University, 66123 Saarbruecken, Germany
}
\author{Anil Kumar Dasanna}
\affiliation{Theoretical Physics of Living Matter, Institute of Biological Information Processing and Institute for Advanced Simulation, Forschungszentrum Jülich, 52425 Jülich, Germany}
\author{Thomas John}%
\affiliation{Experimental Physics, Saarland University, 66123 Saarbruecken, Germany
}
\author{Gerhard Gompper}
\affiliation{Theoretical Physics of Living Matter, Institute of Biological Information Processing and Institute for Advanced Simulation, Forschungszentrum Jülich, 52425 Jülich, Germany}
\author{Lars Kaestner}
\affiliation{Experimental Physics, Saarland University, 66123 Saarbruecken, Germany
}
\affiliation{Theoretical Medicine and Biosciences, Saarland University, 66424 Homburg, Germany}
\author{Dmitry A. Fedosov}
\affiliation{Theoretical Physics of Living Matter, Institute of Biological Information Processing and Institute for Advanced Simulation, Forschungszentrum Jülich, 52425 Jülich, Germany}
\author{Christian Wagner}
\affiliation{Experimental Physics, Saarland University, 66123 Saarbruecken, Germany
}
\affiliation{Physics and Materials Science Research Unit, University of Luxembourg, Luxembourg City, Luxembourg}

\date{\today}

\begin{abstract}
This document gathers Supplemental Figures and description of Supplemental Movies from the aforementioned article. 
\end{abstract}

\maketitle

\section{Details on the fitting process}

While Eq.~(5) does not have an analytical solution, one can show that its inverse function does. One can then write
\begin{equation}
    f(h)=f(h_0)-\frac{\Delta \rho g a^2\phi_m^3}{\kappa_0 \eta\phi_0 h_0} t,
    \label{ShortMain}
\end{equation}
where the function $f$ is defined as
\begin{eqnarray}
    f(x)&=\log{\left(\phi_m x -\phi_0 h_0\right)} +\nonumber\\&\frac{\phi_0 h_0\bigl(\phi_0 h_0 (3-\phi_m)-2(2-\phi_m)\phi_m x\bigr)}{2\left(\phi_0 h_0 -\phi_m x\right)^2}.
    \label{DetailMain}
\end{eqnarray}
In the extreme case where $\phi_m=1$, this function $f$ reduces to
the simpler form
\begin{equation}
    f(x)=\log\left(x-\phi_0 h_0\right)+\frac{\phi_0 h_0}{\phi_0h_0-x}.
\end{equation}
Combination of Eqs.~(\ref{ShortMain}) and (\ref{DetailMain}) can then be used to compute numerically $h(t)$. In order to compare this theoretical curve with experimental data, one has still to evaluate the various physical parameters of the system. We used $g=9.81\,\mathrm{ms^{-2}}$ and measured on average $\Delta \rho\approx80\,\mathrm{kg/m^3}$ in agreement with previous ranges obtained in literature \cite{norouzi2017sorting,trudnowski1974specific}. We considered the viscosity of the liquid to be the average plasma viscosity $\eta=1.2\,\mathrm{mPa\,s}$ \cite{kesmarky2008plasma}. The height of the container $h_0$ in our case is $40\,\text{mm}$. We considered the typical size $a=R r_{\text{RBC}}$, where $r_{\text{RBC}}=4\,$\textmu{m} is the average disk radius of an erythrocyte and $R$ is a parameter reflecting the ratio between the size of the erythrocytes and the characteristic size of the channels (initially unknown). The ratio $\gamma=\frac{\kappa_0}{R^2}$ can then be used as a dimensionless fit parameter related to the characteristic time of the system. The other unknown parameter $\phi_m$, that determines the end of the sedimentation curve can be used as a second fit parameter. Eventually, colloidal gel is known to experience a delayed collapse, whose origin is still under debate although it likely arises from the time required for the cracks to initially form in the gel \cite{teece2014gels,buscall2009towards,gopalakrishnan2006linking,padmanabhan2018gravitational,bartlett2012sudden}. Such a delay time was also previously observed for blood and introduced in a previous empirical model \cite{hung1994erythrocyte}. We then actually introduced this delay time $t_0$ as a third and last parameter, implying
\begin{eqnarray}
    h(t)=
    \begin{cases}
    h_0,&\quad\text{if}~t<t_0\\
    f^{-1}\left(f(h_0)-\frac{G\phi_m^3}{\gamma} (t-t_0)\right)&\quad\text{if}~t\geq t_0
    \end{cases},
    \label{FinalEq}
\end{eqnarray}
where $G=\frac{\Delta \rho g r_{\text{RBC}}^2}{\eta \phi_0 h_0}$, the fit parameter $\phi_m$ being also included in the function $f$.

\section{Legends of Supplemental Movies and Supplemental Figures}
\setcounter{figure}{0}
\renewcommand{\figurename}{Supplemental Movie}
\renewcommand{\thefigure}{S\arabic{figure}}

Supplementray Movies are available at \url{https://drive.google.com/file/d/1G-J8mzo7tlW6YxxNIYlbH-GSBNoZXlYt/view?usp=sharing}, as long as the manuscirpt isn't published in a peer-review journal.

\begin{figure}[!htbp]
    \caption{\textbf{Separated aggregates of erythrocytes during sedimentation} (Hem20StartSpedUpX30.avi) Mesoscopic observation of the sedimentation of erythrocytes, with initial hematocrit of $\phi=0.2$. One can see distinct aggregates falling down, with strong hydrodynamics interactions with uplflowing plasma. Movie recorded with a horizontal microscope and a 8X objective. The full blood is in a container of approximately 150\,\textmu m of thickness and illuminated with a blue led. The total width of the observed area is approximately 5 mm. The movie is 30 times faster than real time.
}
    \label{SuppMov3}
\end{figure}

\begin{figure}[!htbp]
    \caption{\textbf{Gel structure of erythrocytes during sedimentation} (Hem45SpedUp150X.avi) Mesoscopic observation of the sedimentation of erythrocytes, with initial hematocrit of $\phi=0.45$. One can see the apparition of cracks within the packed erythrocytes, which allows the plasma to flow upward. Several eruptions can be seen at the macroscopic interface between packed erythrocytes and the free plasma. The position of the sample is periodically moved to follow the interface. Movie recorded with a horizontal microscope and a 8X objective. The full blood is in a container of approximately 150\,\textmu m of thickness and illuminated with a blue led. The total width of the observed area is approximately 1 cm. The movie is 150 times faster than real time.
}
    \label{SuppMov2}
\end{figure}

\begin{figure}[!htbp]
    \caption{\textbf{Close-up movie of an eruption} (CloseUpEruptionHem45.avi) Observation of an eruption : local uprising of erythrocytes to let the plasma flow upward. Movie recorded with a horizontal microscope and a 20X objective, at the macroscopic interface between free plasma and packed erythrocytes. The initial hematocrit has been fixed at $\phi=0.45$, the full blood is in a container of approximately 150\,\textmu m of thickness and illuminated with a blue led. The blue light being absorbed by hemoglobin, this ensures a better contrast between the plasma and the erythrocytes. The total width of the observed area is approximately 530 \textmu m. The movie is in real time.
}
    \label{SuppMov1}
\end{figure}

\setcounter{figure}{0}
\renewcommand{\figurename}{Supplemental Figure}

\begin{figure*}[!htbp]
    \includegraphics[width=1.0\linewidth]{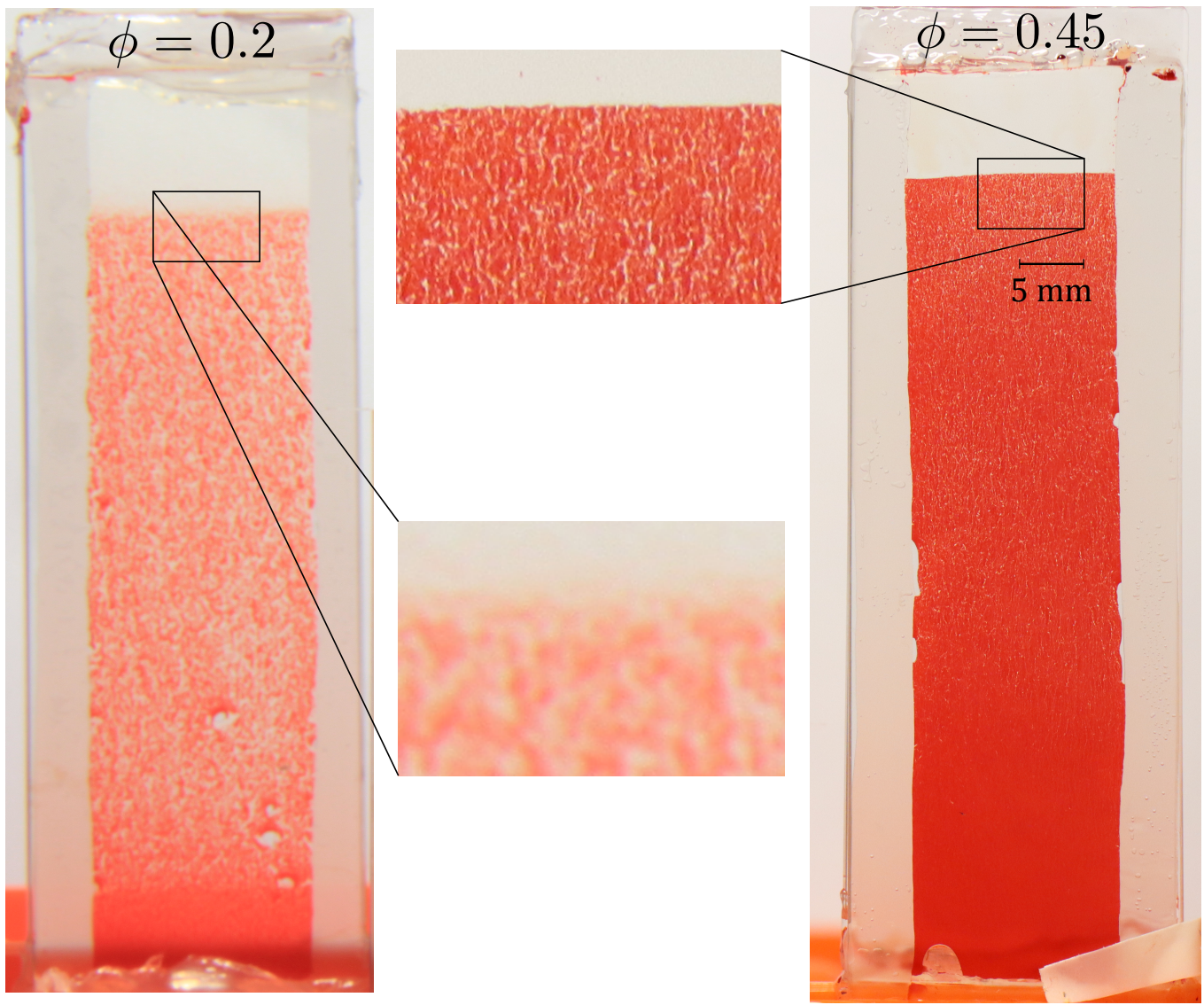}
    \caption{\textbf{Various regimes of erythrocytes sedimentation} Macroscopic observation of the sedimentation of erythrocytes, with initial hematocrit of $\phi=0.2$ (left) and $\phi=0.45$ (right). The full blood is between two glass plates separated by a parafin layer of approximately 150\,\textmu m of thickness, which allows one to see the inherent structure with naked eyes or a classical camera (see also Supplemental Movies S1 and S2 for microscopic zoom level).
}
    \label{SuppFig1}
\end{figure*}

\begin{figure*}[!htbp]
    \includegraphics[width=1.0\linewidth]{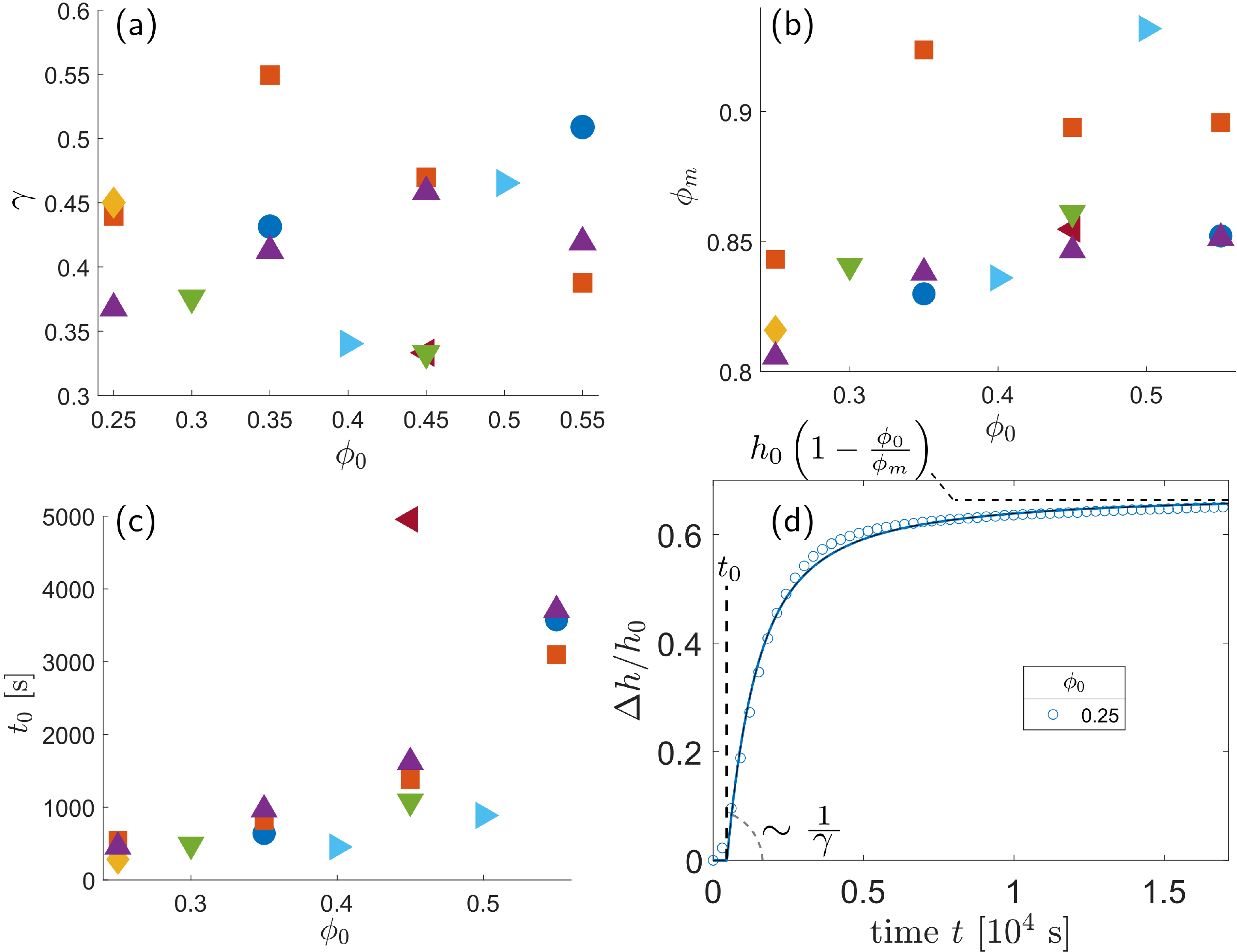}
    \caption{\textbf{Complete range of measured parameters for healthy donors}  Complete range of fit parameters obtained for the various healthy samples considered, as a function of the initial hematocrit $\phi$. Different symbols denote samples from different blood drawings. a) Parameter $\gamma$, dimensionless time of the collapse, modifying the initial slope. b) Parameter $\phi_m$, final volume fraction reached by the erythrocytes. c) Parameter $t_0$, delay time of the collapse. While the previous parameters don't show an obvious trend as a function $\phi$, this $t_0$ time has a tendency to increase with the initial volume fraction $\phi$. d) Schematic representation of the influence of each fit parameter.
}
    \label{SuppFig2}
\end{figure*}

\bibliography{apssampSuppMat}